\documentclass[aps,prb,twocolumn,showpacs,groupedaddress]{revtex4}
\usepackage{graphicx}
\begin{document}


\title{Two superconducting gaps in LaFeAsO$_{0.92}$F$_{0.08}$ revealed by $^{75}$As nuclear quadrupole resonance}

\author{S. Kawasaki$^1$}
\author{K. Shimada$^1$}
\author{G. F. Chen$^2$}
\author{J. L. Luo$^2$}
\author{N. L. Wang$^2$}
\author{Guo-qing Zheng$^1$}

\affiliation{$^1$Department of Physics, Okayama University, Okayama 700-8530, Japan} 
\affiliation{$^2$Institute of Physics, Chinese Academy of Sciences, Beijing 100190, China}





\begin{abstract}
We report $^{75}$As nuclear quadrupole resonance (NQR) studies on superconducting oxypnictide LaFeAsO$_{0.92}$F$_{0.08}$ ($T_{\rm c}$ = 23 K).  The temperature dependence of the spin lattice relaxation rate (1/$T_1$) decreases below $T_{\rm c}$ without a coherence (Hebel-Slichter) peak and shows a temperature dependence that is not simple power-law nor exponential. We show that the result can be understood in terms of two superconducting gaps of either $d$- or ${\pm}s$-wave symmetry, with the larger gap $\Delta_1\sim 4 k_{\rm B}T_{\rm c}$ and the smaller one $\Delta_2 \sim 1.5 k_{\rm B}T_{\rm c}$. Our result suggests that the multiple-gaps feature is universal in the oxypnictides superconductors, which is probably associated with the multiple electronic bands structure in this new class of materials. We also find that 1/$T_1T$ above $T_{\rm c}$ increases with decreasing temperature, which suggests  weak magnetic fluctuations in the normal state. 
\end{abstract}


\maketitle


The discovery of superconductivity in LaFeAsO$_{1-x}$F$_x$ at $T_{\rm c}$ = 26 K \cite{Kamihara}, followed by that in ReFeAsO$_{1-x}$F$_x$ (Re: Ce, Pr, Nd, Sm) \cite{Chen,Ren,Ren2,Chen2,Ren3} with superconducting transition temperature ($T_{\rm c}$) as high as 55 K, has attracted much attention. 
There are many issues remaining to be resolved regarding the physical properties and the pairing mechanism.  
To this end, it is important to unravel the nature of the superconducting energy gap.  Nuclear magnetic 
resonance (NMR) measurements in PrFeAsO$_{0.89}$F$_{0.11}$ ($T_{\rm c}$ = 45 K) found that the Knight shift with magnetic field parallel to the $ab$-plane decreases below $T_{\rm c}$ and goes to zero, which is a strong evidence for spin-singlet pairing in the superconducting state \cite{Matano}. 
The  Andreev reflection measurement on SmFeAsO$_{0.85}$F$_{0.15}$ suggested a  BCS-like gap\cite{ChenSm}.  On the other hand, 
both the Knight shift and the spin-lattice relaxation rate (1/$T_1$) in PrFeAsO$_{0.89}$F$_{0.11}$ suggest that there are two gaps opening below $T_{\rm c}$, with the larger gap $\Delta_1(T=0)$ = 3.5 $k_{\rm B}T_{\rm c}$ and the smaller gap $\Delta_2(T=0)$ = 1.1$k_{\rm B}T_{\rm c}$ in that compound. This observation was echoed by other measurements such as torque magnetometry measurements \cite{Weyenech,Balicas}, point contact tunneling \cite{Samuely,Wen}, and angle-resolved photoemission spectroscopy \cite{Ding}. Thus, the experimental results suggest that the superconducting gap structure resembles that of MgB$_{2}$ \cite{Choi}. These measurements, however, were performed for materials with $T_{\rm c}$ higher than 38 K. It is therefore  natural to ask if the multiple gap feature is unique to the materials with high $T_{\rm c}$ or it is universal in this class of materials.

In this paper, we report a nuclear quadrupole resonance (NQR) study on LaFeAsO$_{0.92}$F$_{0.08}$ ($T_{\rm c}$ = 23 K). NQR has several advantages over NMR. For example, it is performed at zero magnetic field, which gets rids of the complexity due to vortex which makes the $T_1$ at low-$T$ ambiguous as pointed out by Grafe {\it et al} \cite{Grafe}. Also, $^{75}$As nucleus has a nuclear spin $I$ = 3/2, which results in only one transition line of NQR and the recovery curve of the nuclear magnetization is simple and thus, the evaluation of $T_1$ is straightforward and unambiguous.
Our careful and precise measurement shows that $1/T_1$ below $T_{\rm c}$ decreases without a coherence (Hebel-Slichter) peak, but does not follow  simple power-law nor exponential variation as found in high-$T_{\rm c}$ cuprates or MgB$_2$; instead it shows a step-wise temperature variation. The result can be fitted with two gaps of either $d$- or ${\pm}s$-wave symmetry with $\Delta_1(0) \sim$ 4 $k_{\rm B}T_{\rm c}$ and the smaller gap $\Delta_2(0)\sim$1.5 $k_{\rm B}T_{\rm c}$. We will compare our result with those reported earlier for LaFeAsO$_{1-x}$F$_x$ with different carrier concentrations (F content) \cite{Nakai,Grafe,Mukuda,Imai}.

The poly-crystals of LaFeAsO$_{0.92}$F$_{0.08}$ was synthesized by solid state reaction method \cite{ChenPRL}. 
Fine powders of LaAs (pre-synthesized by La pieces and As powder), Fe, Fe$_2$O$_3$, LaF$_3$ with purities better than 99.99\% were mixed together according to the stoichiometric ratio, then ground thoroughly and pressed into small pellets. 
The pellets were sintered at temperature of 1150 $^{\rm o}$C for 50 hours. 
The sample was characterized by powder X-ray diffraction (XRD) method with Cu-$K\alpha$ radiation to be a tetragonal 
ZrCuSiAs-type structure ($P$4$/nmm$ space group). For NQR measurements, the pellet was crushed into coarse powders. 
AC susceptibility measurement using the NQR coil indicates that $T_{\rm c}$ for the powdered sample is 23 K. 
NQR measurements were carried out by using a phase-coherent spectrometer.  
The spin-lattice relaxation rate ($1/T_1$) was measured by using a single saturation pulse.  To avoid possible heating effect due to the rf pulse, we used small amplitude rf saturation pulse to obtain $T_1$ below $T_{\rm c}$. 

\begin{figure}[h]
\includegraphics[width=6.5cm]{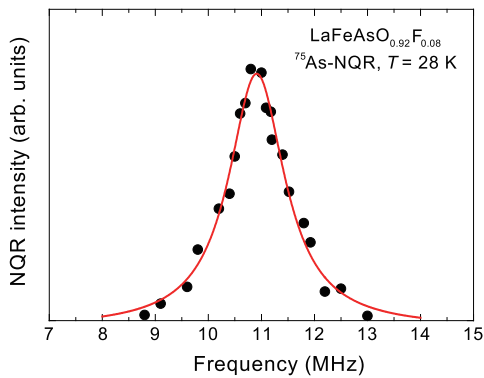}
\caption{\label{fig1} (color online)
$^{75}$As-NQR spectrum at  $T$ = 28 K (above $T_{\rm c}$). Solid curve is a Lorentzian fitting which gives a full width at half maximum (FWHM)  $\sim$ 1.2 MHz.}
\end{figure}

Figure 1 shows the $^{75}$As-NQR spectrum in LaFeAsO$_{0.92}$F$_{0.08}$ ($T_{\rm c}$ = 23 K). The clear single peak structure is observed and it can be fitted by a single Lorentzian curve, indicating that there are no other phases in present sample. The NQR frequency $\nu _{\rm Q}$ is $\sim$ 10.9 MHz, which is in good agreement with earlier reports \cite{Nakai,Grafe,Mukuda}, but smaller than  $\nu _{\rm Q}$ = 12.0 MHz in the higher-$T_{\rm c}$ compound PrFeAsO$_{0.89}$F$_{0.11}$ ($T_{\rm c}$ = 45 K) \cite{Matano}.

Figure 2 shows the representative decay curves of the nuclear magnetization above and below $T_{\rm c}$ = 23 K. Since $^{75}$As nucleus has nuclear spin $I$ = 3/2, the nuclear magnetization for NQR transition $\pm$1/2 $\leftrightarrow$ $\pm$3/2 of $^{75}$As nucleus is given by single exponential 1-$M(t)$/$M_0$ = $\exp$(-3$t/T_1$), where $M_0$ and $M(t)$ are the nuclear magnetization in the thermal equilibrium and at a time $t$ after saturating pulse, respectively. As seen in Fig. 2, $T_1$ is of a single component even well below $T_{\rm c}$.

\begin{figure}[h]
\includegraphics[width=7cm]{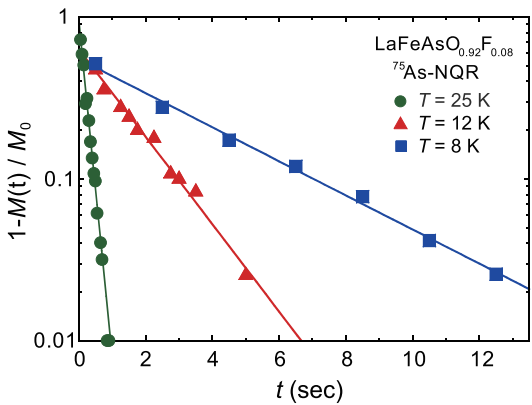}
\caption{\label{fig2} (color online) Decay curves of the nuclear magnetization observed at $T$ = 25, 12, and 8 K, respectively, which indicate a single-component of $T_1$ and allows us an adequate determination of the values of $T_1$. Solid curves are the theoretical fittings to obtain $T_1$.(see text)}
\end{figure}

Figure 3 shows the temperature dependence of $^{75}$($1/T_1$) for LaFeAsO$_{0.92}$F$_{0.08}$. The onset of superconducting transition is clearly observed in temperature dependence of $1/T_1$ below $T_{\rm c}$ = 23 K.
$1/T_1$ decreases with no coherence peak just below $T_{\rm c}$, which is in contrast with conventional BCS superconductor.
Furthermore, the temperature dependence below $T_{\rm c}$ is not a simple power law ($1/T_1$ $\propto$ $T^3$ or $T^5$) as seen in heavy fermion compounds \cite{Zheng,Katayama,Kawasaki} or high-$T_{\rm c}$ cuprates \cite{Asayama}, nor exponential as seen in conventional BCS superconductors \cite{Masuda}.

The most peculiar feature is that, $1/T_1$ shows a step-wise decrease below $T_{\rm c}$. Namely, the steep drop of $1/T_1$ just below $T_{\rm c}$ is gradually replaced by a slower change below $T\sim$ 10 K, then followed by a still steeper drop below. Such unusual decreases of $1/T_1$ leaves  a broad hump-like feature around 8 K. 
 This behavior is clearly different from  the case of usual superconductors that have a single superconducting gap. 
 It should be emphasized that this uncommon temperature variation is not due to sample inhomogeneity, which would result in a two-component $T_1$ below $T_{\rm c}$. We find that $T_1$ is of single component throughout the whole temperature range, as exemplified in Fig. 2. Thus, the two-steps feature in the $T$-dependence of $1/T_1$ appears to be universal in ReFeAsO$_{1-x}$F$_x$, as first found in  PrFeAsO$_{0.89}$F$_{0.11}$ \cite{Matano}.

\begin{figure}[h]
\includegraphics[width=6.5cm]{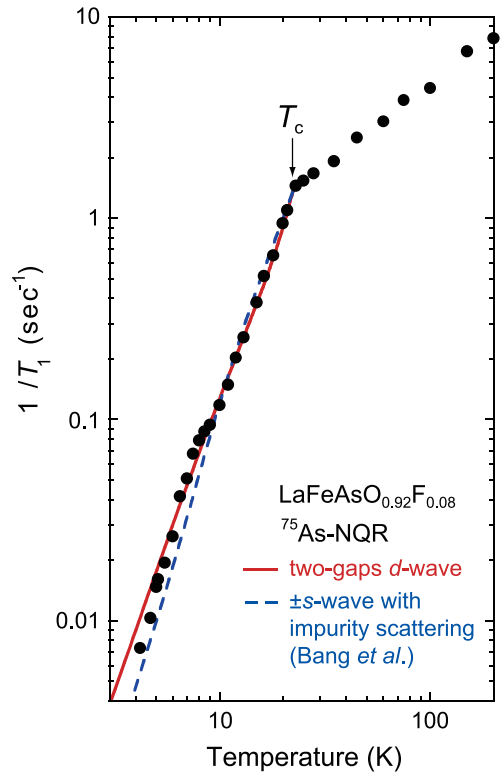}
\caption{\label{fig3} (color online)
The temperature dependence of $^{75}$($1/T_1$) in LaFeAsO$_{0.92}$F$_{0.08}$ measured at zero magnetic field. Experimental error is within the size of the symbols.
The solid curve is a two gap fit assuming a $d$-wave symmetry with  parameters, $\Delta_1(0) = 4.2 k_{\rm B}T_{\rm c}$, $\Delta_2(0) = 1.6 k_{\rm B}T_{\rm c}$,  and  $\alpha = 0.6$ (see text). The dotted curve is a simulation assuming two $s$-wave gaps that change signs with impurity scattering. The parameters are $\Delta_1(0) = 3.75 k_{\rm B}T_{\rm c}$, $\Delta_2(0) = 1.5 k_{\rm B}T_{\rm c}$, and  $\alpha = 0.38$.\cite{Bang,Bang2} The solid arrow  indicates $T_{\rm c}$.}
\end{figure}

In the superconducting state, we find that two-gap models can explain the step-wise temperature variation of $1/T_1$. 
The underlying physics is that the superconducting density of states (DOS) just below $T_{\rm c}$ is dominantly governed by a larger gap while at low temperature it starts to realize the existence of a smaller gap, resulting in another drop $1/T_1$ below 8 K.  
Here, the relaxation rate below $T_{\rm c}$ ($1/T_{1s}$) can be expressed as,  

\begin{widetext}
\begin{eqnarray*}
\frac{T_{1N}}{T_{1s}}= \frac{2}{k_BT} \int \int \left(1+\frac{\Delta^2}{EE'}\right) N_s(E)N_s(E') \left[ 1-f(E') \right] \delta(E-E')dEdE'
\end{eqnarray*}
\end{widetext}

Where 1+$\frac{\Delta^2}{EE'}$ is coherence factor and  $Ns=\frac{E}{\sqrt{E^2- \Delta^2}}$ is the DOS in the superconducting state, and $f(E)$ is the Fermi distribution function. 
In the $d$-wave model with two gaps, $N_{s,i}(E)$ = $N_{s,i}$$\frac{E}{\sqrt{E^2-\Delta_i^2}}$, $1/T_{1s}$ is written as,

\begin{widetext}
\begin{eqnarray*}
\frac{T_{1N}}{T_{1s}}= \sum_{i=1,2}{  \frac{2}{k_BT} \int \int N_{s,i}(E)N_{s,i}(E') \left[ 1-f(E') \right] \delta(E-E')dEdE' }
\end{eqnarray*}
\end{widetext}

and

\begin{eqnarray*}
\alpha = \frac{N_{s,1}}{N_{s,1}+N_{s,2}}
\end{eqnarray*}

By assuming two gaps of $d$-wave symmetry $\Delta(\phi) = \Delta_0 \cos(2\phi)$ with the mean-field temperature dependence, we find that  $\Delta_1(0) = 4.2 k_{\rm B}T_{\rm c}$, $\Delta_2(0) = 1.6 k_{\rm B}T_{\rm c}$  and  $\alpha = 0.6$ can fit the data reasonably well as shown by the solid curve in Fig.3. These values of superconducting gap and $\alpha$ are slightly larger than that obtained in PrFeAsO$_{0.89}$F$_{0.11}$ ($\Delta_1(0)$ = 3.5 $k_{\rm B}T_{\rm c}$, $\Delta_2(0)$ = 1.1$k_{\rm B}T_{\rm c}$, and $\alpha$ = 0.4).\cite{Matano}  The difference may arise from difference in detail of the band structure.

We also compared our experimental results with the theoretical calculation assuming the so-called ${\pm}s$-gap symmetry.
Several theoretical works proposed $s$-wave gap opening on two different Fermi surfaces (with respectively hole and electron characters),  but with opposite sign (${\pm}s$-gap) \cite{Mazin,Kuroki,Lee}. It has been shown that such ${\pm}s$-gap symmetry, with a nodal plane that does not intersect the Fermi surface, gives a higher $T_{\rm c}$ over $d$-wave \cite{Tesanovic,Hu,Bang}. Calculations have shown that magnetic scattering and/or impurity scattering between the two different bands can reduce the coherence peak just below $T_{\rm c}$ \cite{Chubukov,Parker,Hu,Bang,Bang2,Hayashi}. The dotted curve in Fig. 3 is the result by Bang\cite{Bang,Bang2} with the fitting parameters, $\Delta_1(0) = 3.75 k_{\rm B}T_{\rm c}$ with -$s$ symmetry, $\Delta_2(0) = 1.5 k_{\rm B}T_{\rm c}$ with +$s$ symmetry, $\alpha = 0.38$, and $\Gamma$/$\Delta_1$ = 0.04. Here, $\Gamma$ = $n_{\rm imp}$/$\pi$$N_{\rm tot}$, where $n_{\rm imp}$ is the non-magnetic impurity concentration and $N_{\rm tot}$ is the total density of state of the two bands\cite{Bang,Bang2}. As seen in the figure, such ${\pm}s$-gap model also reproduces the overall trend of the experimental results. More work is needed  to distinguish between the $d$- and ${\pm}s$-gap symmetry.

Such two-gaps feature was first pointed out  by the Knight shift and $1/T_1$ measurements in PrFeAsO$_{0.89}$F$_{0.11}$ \cite{Matano}.  
In Fig. 4, we compare the present data with PrFeAsO$_{0.89}$F$_{0.11}$ and with that reported by Grafe $et$ $al$ \cite{Grafe} in LaFeAsO$_{1-x}$F$_x$ with higher F-concentration. The data below $T_{\rm c}$ for PrFeAsO$_{0.89}$F$_{0.11}$ and LaFeAsO$_{0.92}$F$_{0.08}$ are quite similar. Even though the normal state data between ours and Grafe $et$ $al$ are quite different as discussed later in detail, the data below $T_{\rm c}$ are quite similar, although the authors of Ref.\cite{Grafe} analyzed their data in terms of a single gap. Thus, we suggest that the multiple gaps feature is an intrinsic property that is universal to the whole family of Fe-based pnictide superconductors.   

\begin{figure}[h]
\includegraphics[width=7cm]{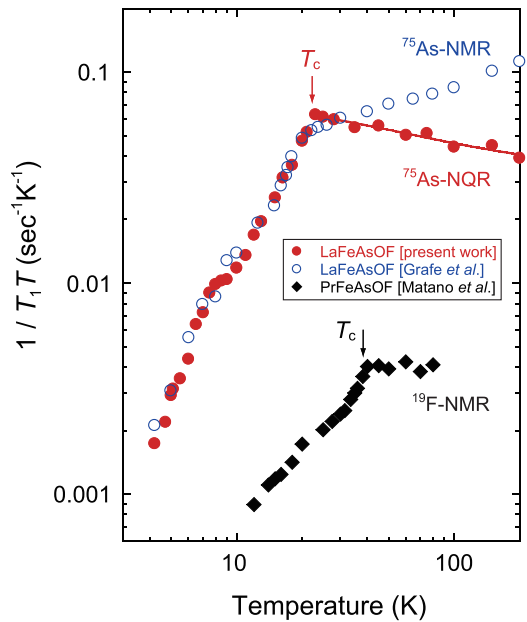}
\caption{\label{fig4} (color online) Temperature dependence of 1/$T_1T$ for LaFeAsO$_{0.92}$F$_{0.08}$ together with LaFeAsO$_{0.9}$F$_{0.1}$\cite{Grafe} and PrFeAsO$_{\rm 0.89}$F$_{0.11}$\cite{Matano} referred from literatures. Arrows indicate $T_{\rm c}$. The solid curve indicates relation, 1/$T_1T$ = 0.03 + $1.8/(T+39)$ in the unit of $\rm{Sec^{-1}K^{-1}}$. (see text)  }
\end{figure}

Finally, we discuss normal-state property in LaFeAsO$_{0.92}$F$_{0.08}$. As seen in Fig.4, the quantity of $1/T_1T$ in our sample does not show a reduction above $T_{\rm c}$ which was ascribed to a possible pseudogap behavior observed in LaFeAsO$_{1-x}$F$_x$ \cite{Imai,Nakai,Grafe}. Instead, it increases with decreasing $T$. This may be due to the fact that our sample has a lower concentration of F than others \cite{Imai,Nakai,Grafe}. The temperature dependence of 1/$T_1T$ above $T_{\rm c}$ in LaFeAsO$_{0.92}$F$_{0.08}$ is well fitted by the relation for a weakly antiferromagnetically-correlated metal, 1/$T_1T$ = $C/(T+\theta)$ + const. \cite{Moriya}. Here, the first term described the contribution from the   antiferromagnetic wave vector $Q$, and the second term describes the contribution from the density of states at the Fermi level.   As shown by the  solid curve in Fig. 4, the temperature dependence of $^{75}$(1/$T_1T$) for  LaFeAsO$_{0.92}$F$_{0.08}$ is well represented by this model with $\theta$ $\sim$ 39 K.  Qualitatively similar behavior was also seen in  underdoped high-$T_{\rm c}$ cuprates \cite{Ohsugi} or cobaltate superconductors \cite{Fujimoto}, but $1/T_1T$ increases much more steeply there with smaller $\theta$. The result  indicates that the spin correlations are much weaker in the present case.

In conclusion, we have presented the detailed NQR measurements on oxypnictide superconductor LaFeAsO$_{0.92}$F$_{0.08}$ ($T_{\rm c}$ = 23 K). The temperature dependence of $1/T_1$ shows no coherence peak just below $T_{\rm c}$ and it decreases with a small hump at around $T$ $\sim$ 0.4 $T_{\rm c}$, suggesting  multiple-gaps  of the superconductivity. Assuming two gaps  of either $d$- or $\pm$$s$-wave can reproduce whole temperature dependence of $1/T_1$ qualitatively.  We suggest that the multi-gap superconductivity is an intrinsic and universal feature  in this class of materials which is likely originated from their multiple electronic band structure. We also find that the normal state of LaFeAsO$_{0.92}$F$_{0.08}$ is governed by the weak magnetic fluctuations which are commonly observed in under doped high-$T_{\rm c}$ cuprates or superconducting cobaltate. It is the future work to clarify the relationship between the magnetic fluctuations and the high $T_{\rm c}$ superconductivity in these compounds.

We thank Prof. Yunkyu Bang for fruitful discussions and for providing the result of theoretical calculation. We also thank Z. Li and T. Oka for contributions to this work.  
This work was supported in part by research grants from MEXT and JSPS, Japan, NSFC of China, 
and the ITSNEM program of CAS.


\end{document}